\documentclass[10pt]{iopart}
\usepackage{graphicx}
\usepackage[normalem]{ulem}

\usepackage{color}

\begin{document}

\title[N Rougemaille \textit{et al.} 2D Materials (2019) \href{https://doi.org/10.1088/2053-1583/ab111e}{DOI:10.1088/2053-1583/ab111e}]{Confined step-flow growth of Cu intercalated between graphene and a Ru(0001) surface}

\author{Nicolas Rougemaille$^1$, Sergio Vlaic$^2$, Lucia Aballe$^3$, Michael Foerster$^3$, Johann Coraux$^1$}
\address{$^1$ Univ. Grenoble Alpes, CNRS, Grenoble INP, Institut NEEL, 38000 Grenoble, France}
\address{$^2$ Laboratoire de Physique et d'\'{E}tude des Mat\'{e}riaux, ESPCI Paris, PSL Research University, CNRS, Sorbonne Universit\'{e}s, UPMC Univ Paris 06, 75005 Paris, France}
\address{$^3$ ALBA Synchrotron Light Source, 08290 Cerdanyola del Valles, Spain}
\ead{johann.coraux@neel.cnrs.fr}

\begin{abstract}

By comparing the growth of Cu thin films on bare and graphene-covered Ru(0001) surfaces, we demonstrate the role of graphene as a surfactant allowing the formation of flat Cu films.  Low-energy electron microscopy, X-ray photoemission electron microscopy and X-ray absorption spectroscopy reveal that depositing Cu at 580~K leads to distinct behaviors on both types of surfaces. On bare Ru, a  Stranski-Krastanov growth is observed, with first the formation of an atomically flat and monolayer-thick wetting layer, followed by the nucleation of three-dimensional islands. In sharp contrast, when Cu is deposited on a graphene-covered Ru surface under the very same conditions, Cu intercalates below graphene and grows in a step-flow manner: atomically-high growth fronts of intercalated Cu form at the graphene edges, and extend towards the center of the flakes. Our findings suggest potential routes in metal heteroepitaxy for the control of thin film morphology.\\
\textit{Accepted for publication in 2D Materials (2019) \href{https://doi.org/10.1088/2053-1583/ab111e}{DOI:10.1088/2053-1583/ab111e}}
\end{abstract}

%
\vspace{2pc}

%
%

\ioptwocol

\section*{Introduction}

The way thin films grow epitaxially on a single crystal surface strongly depends on the respective surface energies, the interaction strength between the deposited adatoms and the surface, and on the epitaxial stress induced by the substrate. Three primary growth modes are usually considered in thin film epitaxy \cite{Bauer1958,Bauer1986}:

\noindent-- The Frank-van der Merwe growth mode, when adatoms attach preferentially to the surface rather than clustering, leading to a layer-by-layer growth,

\noindent-- The Volmer-Weber growth mode, when adatoms preferentially bond together, resulting in the formation of three-dimensional islands,

\noindent-- The Stranski-Krastanov growth mode, which is an intermediate situation characterized by the formation of three-dimensional islands on top of a two-dimensional layer.

These considerations are thermodynamic arguments, and (strong) deviations from these growth modes occur when growth is governed by kinetic limitations. Other surface morphologies may be obtained in specific cases, for example when surface alloying plays a key role \cite{Rougemaille2007,Hannon2006,Schmid2000,Hwang1996}. Surfactants can also be employed to influence surface energetics and thus control thin film morphology \cite{Egelhoff1989,Copel1990,Rosenfeld1993,Sakai1994,Voigtlander1995,Tanaka1996}. In that context, graphene has been recently identified as an interesting surfactant, being a floating \cite{Wang2014}, covalent and deformable membrane that allows intercalation of adatoms and confined growth under a purely two-dimensional (2D) cover \cite{Vlaic2018}. 

Adatoms intercalated between graphene and the substrate obviously interact with both materials. Thus, besides affecting the film morphology, one might also expect graphene to potentially modify the structural, electronic or magnetic properties of the intercalated layer. For instance, graphene has been shown to be a promising capping layer in magnetic systems, especially to promote perpendicular magnetic anisotropy \cite{Rougemaille2012,Decker2013, Vu2016,Yang2016,Gargiani2017}, a redistribution of the magnetic moment perpendicular to the surface \cite{VoVan2010,Abtew2013,Shick2014,Sipahi2014,Yang2016,Dunaevskii2018}, or chiral magnetic textures \cite{Yang2018,Ajejas2018,Genoni2018}.

Intercalation is usually obtained by post-annealing treatments \cite{Tontegode1991,Dedkov2001,Coraux2012,Sung2014,Vita2014,Pacile2014,Drnec2015}. Although a discussion on the microscopic origin of the intercalation process remains often vague, point defects in graphene have been proposed to be potential channels where matter can penetrate \cite{Coraux2012,AlBalushi2016}. Other studies suggest that besides point defects \cite{Sicot2012,Petrovic2013,Vlaic2014}, graphene edges \cite{Sutter2010,Vlaic2017,Vlaic2018} are also efficient intercalation pathways.

In this work, we investigate the growth of Cu films on a graphene-covered Ru(0001) surface, using low-energy electron microscopy (LEEM) and X-ray photoemission electron microscopy (PEEM), combined with spatially-resolved electron reflectivity and photoelectron yield spectroscopy. This system is interesting as graphene is known to strongly interact with Ru \cite{Sutter2009}, while Cu is a metal of choice to decouple epitaxial graphene \cite{Shikin1998,Shikin1999,Farias2000,Dedkov2001} and to modify its electronic properties \cite{Vita2015,Forti2016}. When it is deposited at 580~K, we find that Cu intercalates underneath graphene and then grows in a step-flow manner. In sharp contrast, on bare Ru(0001) and under the very same conditions, Cu growth is characteristic of a Stranski-Krastanov mode: randomly distributed three-dimensional Cu islands form on a wetting layer. These findings highlight the unique surfactant role of graphene and suggest potential routes for controling the morphology of thin metal films in heteroepitaxy.

\section*{Methods}

\begin{figure}[!hbt]
\centering
\includegraphics[width=7.04cm]{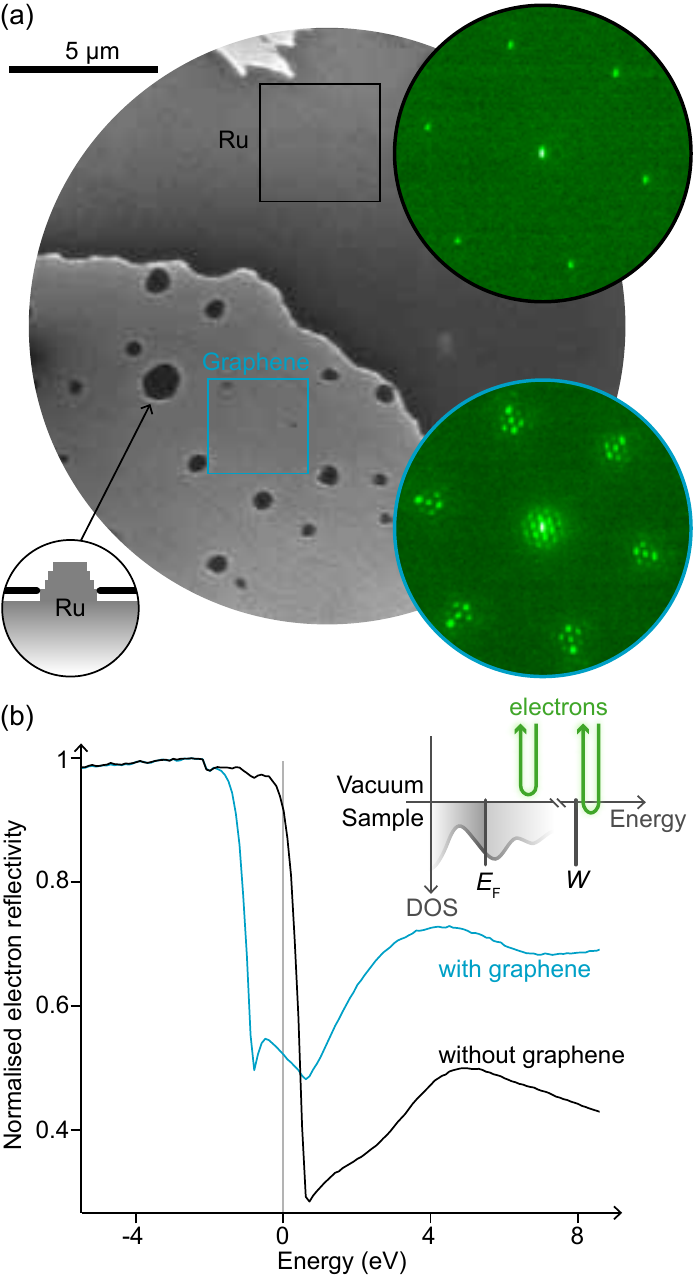}
\caption{Ru(0001) surface partly covered with graphene. (a) LEEM image of single-layer graphene flakes grown on Ru(0001) (electron energy: 2.5~eV). Protruding Ru mesas imprinting holes in graphene are schematized in the bottom-left inset. Micro-diffraction patterns with 150~eV electrons, taken on bare Ru(0001) (black-circled) and on graphene/Ru(0001) (blue-circled), are shown. (b) Electron reflectivity as function of the electron kinetic energy extracted from the two colored frames in (a). The sudden decrease of reflectivity at about 0~eV corresponds to the work function ($W$) of the surface (see inset).}
\label{fig1}
\end{figure}

Experiments were performed under ultrahigh vacuum conditions (base pressure of 4$\times$10$^{-10}$~mbar). A ruthenium single-crystal cut with a (0001) surface was used as a substrate. The surface was prepared by repeated cycles of 820~eV Ar$^+$ ion sputtering and flash-annealing to 1720~K. To reduce the concentration of dissolved carbon close to the surface region, the sample was occasionally annealed at 870~K under a 10$^{-6}$~mbar partial pressure of O$_2$ for 10~min. Just after this treatment, and in any case prior to graphene growth, the sample was flash-annealed to 1510~K. This procedure yielded a clean surface, with no apparent trace of surface contamination or residual graphene flakes as assessed by low-energy electron microscopy (LEEM) and low-energy electron diffraction (LEED).

Graphene growth was then performed by slowly decreasing the substrate temperature in order to promote surface segregation of carbon dissolved into the bulk \cite{Sutter2008,McCarty2009}. We observed that graphene islands start to form below 1120~K, while they rapidly decompose above 1170~K as bulk dissolution of carbon is activated. 
We chose a temperature of about 1100~K for which the typical distance between graphene nuclei is a few to several 10~$\mu$m.
After a $\sim$20-30\% surface coverage was obtained, the sample was quenched to room temperature to stop the growth.

An evaporator was used to deposit 99.99\%-pure Cu on the surface. The deposition rate was determined by LEED and LEEM observations during the completion of a pseudomorphic Cu monolayer on an Ir(111) substrate \cite{Tonner1993}. This rate is about 0.2~equivalent monolayer/min, 1 equivalent monolayer (eq. ML) corresponding to the surface density of a pseudomorphic Cu layer on Ru(0001) (the lattice constant of the (111) surface of bulk Cu is typically 5.5\% larger than the one of the Ru(0001)). 
In the following, we use the "eq. ML" notation to refer to the amount of Cu deposited, and to the "ML" notation to describe the actual Cu thickness.

Copper deposition and graphene growth were performed under ultrahigh vacuum conditions, \textit{in operando}, in front of the objective lens of the Elmitec LEEM-III end-station at the CIRCE beamline of the ALBA Synchrotron \cite{Aballe2015}. Both soft X-ray synchrotron light and low-energy electrons were used as sources for imaging, yielding PEEM and LEEM images, respectively. Electron reflectivity spectroscopy was performed by varying the potential between the electron gun of the microscope and the sample surface (\textit{i.e.} by varying the kinetic energy of the electrons). In addition, X-ray absorption spectroscopy (XAS) in partial electron yield mode was performed by scanning the photon energy across the Cu $L_3$ and $L_2$ absorption edges, while detecting the emitted secondary photoelectrons.

\section*{LEEM observations}

Before addressing the growth of copper, we briefly describe the sample surface after graphene deposition.
In the LEEM image of figure~\ref{fig1}(a), graphene partially covers the Ru(0001) surface (in the imaging conditions used here, graphene appears brighter than bare Ru). We note the presence of holes in the graphene flakes, having sizes of few 100~nm, typically. These graphene holes are presumably a consequence of graphene growth proceeding from the ascending step edges of the Ru(0001) substrate \cite{Sutter2008}, revealing the presence of mesas in the Ru surface (see inset in figure~\ref{fig1}(a)). Thus, the graphene flakes are not continuous, although single crystalline.

The micro-diffraction patterns [figure~\ref{fig1}(a)] on bare and graphene-covered Ru(0001) show the expected (1$\times$1) and surface superstructure \cite{Sutter2008}. The latter corresponds to the coincidence lattice (so-called moir\'{e} pattern) resulting from the lattice mismatch between graphene and Ru(0001) \cite{Marchini2007}. Finally, electron reflectivity (figure~\ref{fig1}(b)) reveals typical features for graphene-free and graphene-covered regions \cite{Sutter2008,Sutter2013}. The onset of total reflection, which corresponds to electrons having a kinetic energy higher than the surface work function ($W$), is decreased by 1.3~eV in presence of graphene, due to charge transfer at the metal-graphene interface \cite{Giovannetti2008}.

This surface, partly covered with graphene, has been used to investigate the growth of Cu at 580~K. Growth modes on bare Ru(0001) and on graphene/Ru(0001) can be then studied simultaneously. The growth temperature was 580~K. Following previous reports \cite{Vickerman1983,Ammer1997}, this temperature was chosen to observe a Stranski-Krastanov growth of Cu on bare Ru(0001). We also used a 700~K growth temperature  to speed up the intercalation process. In these conditions (deposition at 580 and 700~K), the pressure in the system did not exceed the high 10$^{-10}$~mbar-range.

\begin{figure}[!hbt]
\centering
\includegraphics[width=7.4cm]{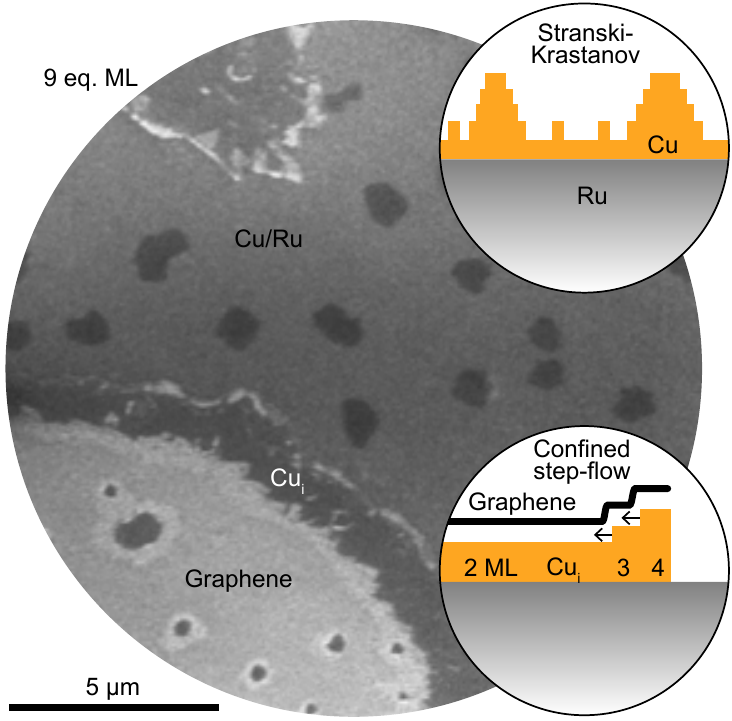}
\caption{Copper on bare Ru(0001) and intercalated (Cu$_\mathrm{i}$) below graphene. LEEM image after the deposition of 9~eq. ML (electron energy: 0.5~eV; same region as the one shown in figure~\ref{fig1}(a)). Underneath graphene, growth proceeds in a step-flow-like fashion, from the edges of the graphene flake. On bare Ru(0001), a characteristic Stranski-Krastanov morphology develops, with a critical thickness for the 2D-3D transition between 1 and 2~ML of Cu.}
\label{fig2}
\end{figure}

\subsection*{Copper growth on bare Ru(0001)}

We first analyse Cu growth on bare Ru based on the LEEM image sequence reported in the supplementary information (SI) figure~S1. After the deposition of about 1.2~eq. ML, the LEEM contrast coming from the atomic step structure of the substrate is recovered (see zoomed-in region in figure~S1), indicating that a complete 1 ML Cu is formed (table~S1, movie~S1). This threshold amount of 1.2~eq. ML is substantially larger than 1~ML. The difference may be related to the calibration of the Cu deposition rate (we estimate the uncertainty to be of the order of 10-15\%). In addition, part of the Cu adatoms might be in a dilute phase on the surface at the 580~K deposition temperature.

Further Cu deposition (see zoomed-in region in figure~S1 for 1.5~eq. ML and movie~S1) leads to a roughening of the surface, suggesting that the growth of the second atomic layer in the form of one-extra-layer-thick islands, separated by few 10~nm distances. Further increasing the amount of deposited Cu never leads to the completion of the second Cu ML.
Instead, a sudden formation of three-dimensional (3D) Cu islands is observed (see zoomed-in region in figure~S1, figure~\ref{fig2}, movie~S1). The 3D islands have a much lower density than the above-discussed 2~ML-thick islands and are typically separated by a few micrometers. We note that the location of these 3D islands is not correlated to the position of the mesas on the Ru(0001) surface.

The low nucleation density of the islands signals a transition from a 2D growth mode to a 3D growth mode: above 1.9~eq. ML, 3D islands coexist with a 2D wetting layer, which is typical for a Stranski-Krastanov growth. The critical thickness for the 2D/3D transition is here between 1 and 2~ML of Cu. Although consistent, this value is different from those reported when deposition is made at slightly lower temperatures of 520~K (3~ML \cite{Ammer1997}) or 540~K (1~ML \cite{Vickerman1983}). The driving force for this transition is presumably not the distinct surface energies of Ru(0001) and Cu(111) ($\sim$3.0~J/m$^2$ and $\sim$1.8~J/m$^2$ respectively \cite{Tyson1977,DeBoer1988}), as the larger value for Ru(0001) should rather promote layer-by-layer growth (Frank-van der Merwe growth). However, the relief of tensile elastic energy in Cu, which results from the 5.5\% lattice mismatch between the two materials, could be a natural driving force for the formation of 3D Cu islands.

Consistent with previous literature, we find that the first Cu atomic layer is in fact pseudomorphic to Ru(0001), hence strongly stretched (see figure~S2). Above this thickness, a variety of stress relief patterns has been documented as a function of post-deposition annealing temperatures \cite{Guenther1995,Zajonz2000,deLaFiguera2001}. The wetting layer that we observe between the 3D Cu islands consists of a full single layer plus a fraction of a double layer, and its diffraction pattern (see figure~S2) indicates (at least) partial stress relief.

\begin{figure}[!hbt]
\centering
\includegraphics[width=8.27cm]{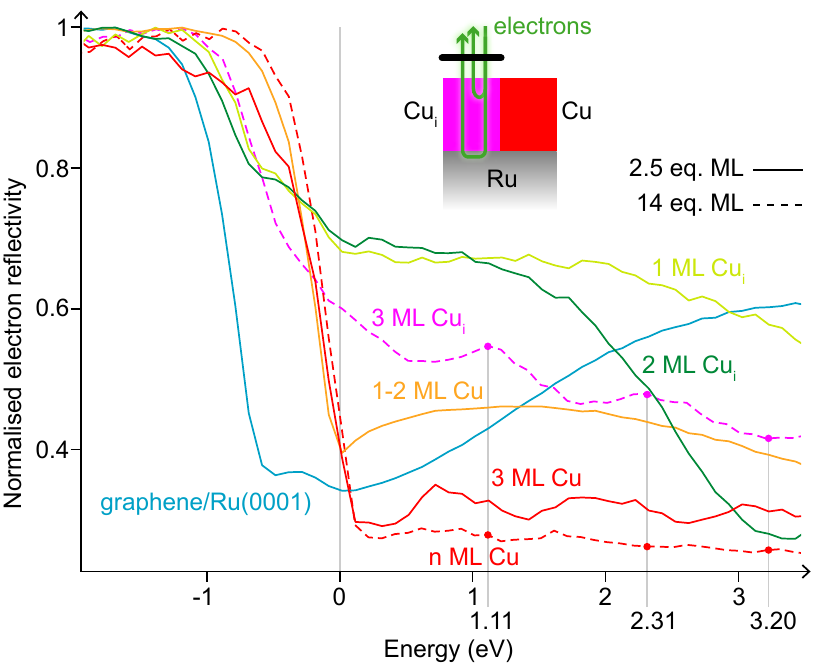}
\caption{Electron reflectivity as function of the kinetic energy after the deposition of 2.5 (full lines) and 14 (dotted lines) eq. ML of Cu, extracted from different regions on the sample surface. Cu and Cu$_i$ stand for Cu on bare Ru and intercalated below graphene, respectively.}
\label{fig3}
\end{figure}

\subsection*{Copper growth on graphene-covered Ru(0001)}

We now describe the real-time growth of Cu on graphene-covered Ru(0001). As Cu is deposited, no visible change in the LEEM contrast is observed on the inner regions of the graphene flakes, contrary to what is found on bare Ru (see figure~S\ref{fig1} for low coverages and table~S1). The only discernable change on graphene-covered regions is the appearance of a rim extending over several 10~nm, located at the edges of the flakes (see figure~\ref{fig2}, figure~S1 for 1.2~eq. ML, and movie~S1). Consistent with other works \cite{Vlaic2018}, this suggests that Cu adatoms have long-range mobility on graphene at the deposition temperature, presumably beyond the few 10~$\mu$m size of the graphene flakes. This range is in agreement with first principle calculations on free-standing graphene \cite{Yazyev2010} but also with related experimental observations \cite{Liu2015}. When reaching the edges of a graphene flake, Cu adatoms then land down on the metal surface and intercalate (see below).

At about the same time (given the typical 2~s time resolution of our measurements) the 3D Cu islands form on bare Ru, a second rim starts to grow at the graphene edges (see zoomed-in region in figure~S1 for 1.9~eq. ML). For higher Cu coverages, both rims grow wider and a third (and even a fourth, hardly discernable in the imaging conditions used here) rim forms (movie~S2). These observations thus reveal a step-flow growth mode from the edges of the graphene flakes, in sharp contrast with what is observed on bare Ru.

\begin{figure*}[!hbt]
\centering
\includegraphics[width=14.6cm]{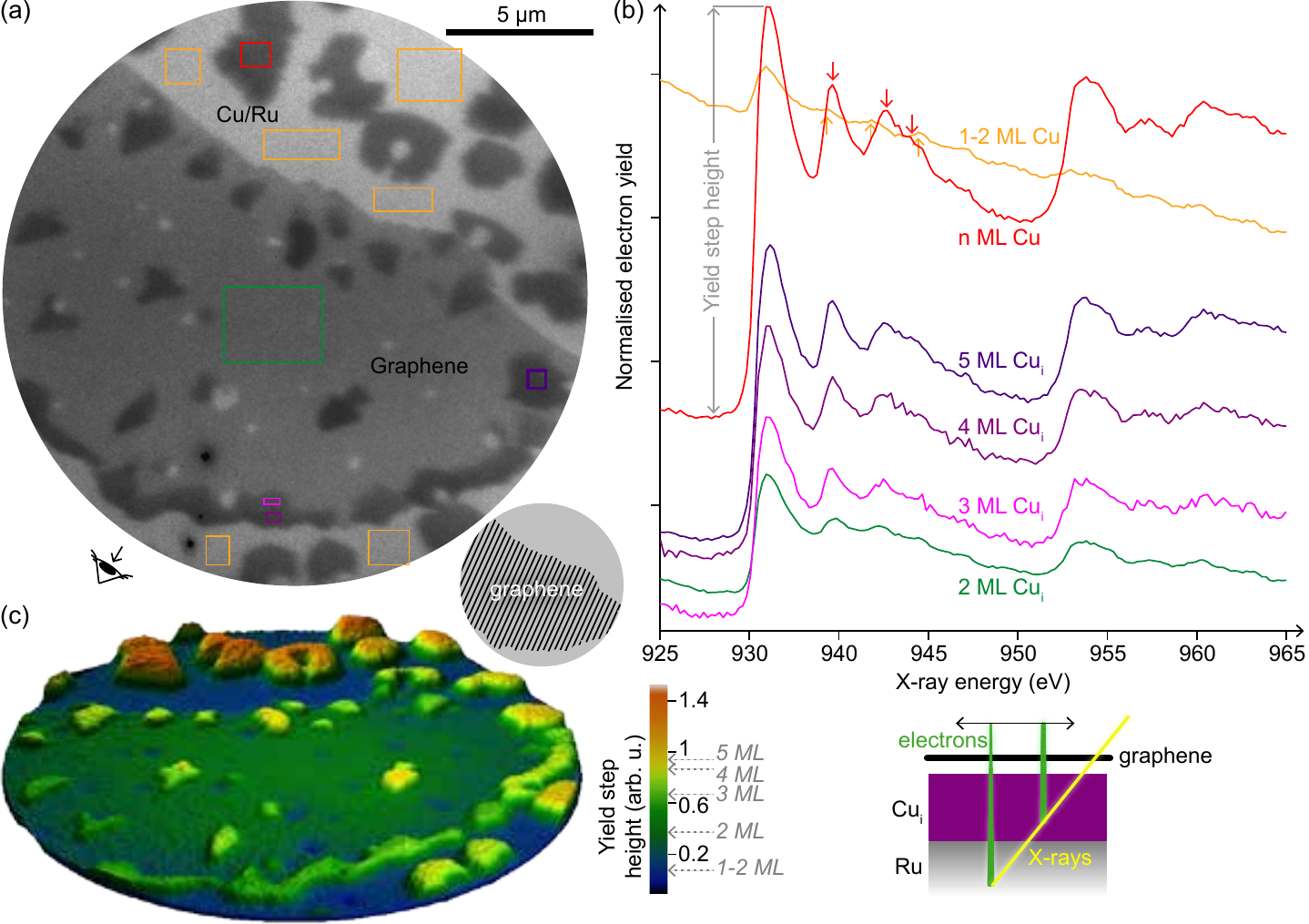}
\caption{Copper thickness-dependent XAS spectra. (a) Pre-edge PEEM image acquired with 927.5~eV X-rays), after the deposition of 14~eq. ML of Cu. In the explored range of X-ray energies, the escape length of the photoelectrons is of the order of 1 nm, significantly smaller than the X-ray penetration depth. A cartoon identifies graphene-covered regions in (a,c). (b) Cu absorption across the $L_{2,3}$ edges extracted from the different regions highlighted in (a). The curves are vertically shifted (reference at 965~eV) for clarity. (c) $L_3$ absorption edge step height (see (b)) extracted point-by-point (with 4 pixel-binning) across the surface, and represented in three dimensions (viewpoint indicated with an arrow an the symbol of an eye in (a)). The height increases as the number of Cu layer increases, although not linearly.}
\label{fig4}
\end{figure*}

The rims that we observe at the graphene edges and that extend into the inner regions of the flakes actually correspond to Cu intercalated between graphene and Ru. To support this claim, we use the fact that the surface work function of graphene-covered Cu is substantially decreased compared to bare Cu. This can be seen in figure~\ref{fig3}, where we plot the electron reflectivity for different regions of the surface. In particular, the green and magenta curves, obtained on the graphene flakes, are shifted by a few 100 meV compared to the orange and red curves measured on Cu/Ru(0001). This reduction of the surface work function for graphene-capped metal surfaces is typical and consistent with other works \cite{Giovannetti2008,Murata2010,Starodub2011}. Interestingly, as we will see below, these Cu rims have a well-defined thickness, ranging from 1 to 4~ML. The fact that Cu can be intercalated underneath a full micron-size graphene flake indicates that Cu diffusion is long-ranged \cite{Liu2015}.

We emphasize here that the growth mode we observe underneath graphene is unusual (see section~4 of the SI) and does not correspond to any of the three modes described in the introduction. In particular, if the intercalation leads to the formation of rather flat Cu films underneath graphene, the growth is not, strictly speaking, layer-by-layer, as a Cu rim with various thicknesses is observed at the edges of the graphene flakes. Interestingly, the growth is not either a conventional step-flow growth as the Ru atomic step structure does not matter much in the intercalation mechanism (although this might not be completely true, see below). What matters however is the step edge of the graphene flakes. Our measurements clearly reveal that the intercalated growth front starts from the graphene edges (of the flakes or the mesas) and advances towards the center of the flake. In that sense, the growth mechanism is step-flow like, but confined under graphene, and essentially independent of the substrate step structure. We also note that the rim with 3-4~ML Cu thickness extends over a fully intercalated 2~ML-thick Cu film. The speed at which the intercalated growth front progresses appears thickness-dependent. This could be related to the fact that the first Cu ML is pseudomorph to the Ru(0001) substrate, while the corresponding tensile strain might be relaxed for thicker films.

\section*{PEEM observations}

Our LEEM observations reveal that Cu is inhomogeneously distributed on the Ru surface, with well-defined thicknesses at the edges of the graphene flakes and thick 3D islands on bare Ru. To confirm this finding, we used chemically-resolved microscopy, combining XAS with PEEM. A typical PEEM image acquired at an energy below the absorption edge is reported in figure~\ref{fig4}(a), after the deposition of $\sim$14~eq. ML of Cu (the last 5~eq. ML have been deposited at 700~K to speed up the intercalation process). The inset identifies the graphene-covered regions.

Recording a series of PEEM images while scanning the photon energy allows extracting local XAS spectra down to the single pixel level (figure~\ref{fig4}(b)). As expected, Cu is found everywhere on the surface, on bare Ru and on graphene-covered regions as well, where Cu is intercalated. Except for the Cu wetting layer on bare Ru, all spectra are rather similar. The larger absorption signal is obtained on the thick Cu islands on graphene-free regions and is similar to the one of bulk Cu \cite{Grioni1992}. For the Cu wetting layer on bare Ru (orange curve in figure~\ref{fig4}(b)), the oscillations after the $L_3$ edge are not located at the same energies as in the other spectra, suggesting a different local environment for Cu atoms. This difference may arise from the strain present in the wetting layer, which is pseudomorphic to Ru(0001).

In figure~\ref{fig4}(c) and figure~S3, the $L_3$ absorption edge height is plotted as a function of the number of Cu layers. Although these two quantities are not directly proportional \cite{Ejima2003}, the step height monotonously increases with the number of Cu layers, so that a map of the Cu thickness can be extracted from the series of PEEM images. Complementary to what we found with LEEM, the rims of intercalated Cu have a 3 to 4~ML thickness, while the remaining of the graphene flake is fully intercalated with 2~ML of Cu. However, thicker intercalated deposits are found. In particular, 5~ML-thick intercalated regions are prominently found at the edges of the graphene flake. Others are also found in the center of the flake, at the location of a graphene hole, which acts as an effective edge.

Noteworthy, intercalated regions thicker than 2~ML are not distributed equally all around the graphene flake shown in figure~\ref{fig4}(a). In particular, intercalation seems to be more efficient at the lower edge in the image.
It could be that the upper edge in the image is located along a bunch of ascending atomic steps of Ru(0001).  
Such graphene edges are known to be strongly bound to the substrate and remain essentially immobile during graphene growth. There, Ru(0001) steps effectively act as walls preventing graphene growth \cite{Sutter2008}. In contrast, the bottom edge in the image would be weakly bound and mobile during graphene growth \cite{Sutter2008}. We expect that these type of mobile edges are preferred pathways for Cu intercalation. Coming back to the discussion on the growth mode of the intercalated Cu film, these PEEM measurements indicate that the Ru atomic step structure has in fact partial influence on the intercalation mechanism. But this influence seems more related to the energy barrier involved to initiate intercalation than the structure of the Ru surface atomic steps.

The variability in the thickness of intercalated Cu at the vicinity of graphene edges may be seen as a tendency to surface roughening. But the dynamics of this roughening is different from the one usually observed in standard layer-by-layer growth.
In fact, in absence of a graphene capping layer, surface roughening is inevitable, even in the case of a Frank-van der Merwe growth, due to the small, but non-zero, probability for two (or more) metal adatoms to meet on a terrasse. Atomic dimers (or trimers, tetramers, etc) are then seeds for the further attachment of adatoms, hence nucleating a new atom-thick layer. The formation of few atom clusters can also occur on these newly created atom-thick islands, and roughening develops accordingly. This cluster mechanism should be less probable when the surface is graphene-covered. This is so because two (or more) adatoms first need to climb up a one-atom-thick island before nucleating a new island. In other words, the fact that growth is confined under graphene and that additional matter can only enter from the graphene edges should significantly lower the probability of roughening. This expected low probability is in fact consistent with the very low density of $>$1~ML (figures~\ref{fig2},S1,S4) or $>2$~ML (figure~\ref{fig4}) islands in the LEEM and PEEM observations. Besides, this probability presumably decreases with the distance from the graphene edges (the density of intercalated adatoms must decrease with the distance from the edge). Contrary to the "standard" case in absence of a capping layer, roughening should be effectively suppressed away from the graphene edges.

\section*{Summary and concluding remarks}

Copper deposition at 580~K on a Ru(0001) surface partly covered with a graphene layer reveals two distinct growth modes. On bare Ru, a Stranski-Krastanov growth is observed and the morphology of the Cu film is characterized by large micron-size, three-dimensional islands, coexisting with a Cu wetting layer of thickness between 1 and 2~ML. On graphene-covered Ru, Cu intercalates at the edges of the graphene flakes and a confined step-flow regime takes place, bringing Cu towards the center of the flakes. The main result of this work is the observation that graphene allows to bypass the formation of three-dimensional islands by modifying the surface energetics and the kinetics of mass transport on the surface, and imposes a step-flow growth. The steps that matter in this confined growth mode are not those of the Ru substrate but those defined by the graphene edges. Interestingly, oxygen has been used also to promote the formation of two-dimensional Cu films on Ru(0001), but following a layer-by-layer growth mode \cite{Wolter1993,Wolter1997}. Here instead, graphene forces a step-flow regime.

A natural question is whether the growth mode we revealed can be implemented, beyond the scale of a few 10~$\mu$m set by the size of the graphene flakes, at a macroscopic length scale. In other words, one might wonder whether confined step-flow growth can be achieved underneath a full graphene layer. Obviously, this question deserves more work, but a prerequisite is that intercalation channels exist for Cu. We note that, unlike the case of Co intercalation between graphene and Ir(111), which comes with a variety of intercalation channels (edges \cite{Vlaic2017,Vlaic2018}, bent graphene regions \cite{Vlaic2014}, point defects \cite{Coraux2012}), here we did not observe other intercalation channels than edges (at 580 and 700~K). Besides, preliminary attempts to intercalate a thin Cu film (6~eq.~ML) deposited at room temperature on top of polycrystalline graphene/Ru(0001) (prepared by catalytic decomposition of ethylene at 870~K; data not shown) and subsequently annealed up to 700~K, were unsuccessful. This is also in contrast with the Co intercalation between polycrystalline graphene and Ir(111) \cite{Coraux2012} or Pt(111) \cite{Ajejas2018}. Point defects in polycrystalline graphene on Ru(0001) do not seem to favor intercalation. Even though we cannot exclude that point defects or other kinds of intercalation channels become active at temperatures beyond the range we explored (580, 700~K), a confined step-flow growth might be induced by generating defects, artificially. These artificial defects could be obtained for example after graphene islands have coalesced, but before a 100\% coverage is achieved, or could be created \textit{a posteriori} by an etching process. After intercalation, a full protection of the surface could be achieved by a second chemical vapor deposition of graphene onto the bare Cu regions (those covering the initially bare Ru(0001) regions, at the location of the graphene defects).

Our work opens new prospects to revisit epitaxial growth with a new ingredient, namely a covalent, deformable graphene layer. By altering both the growth kinetics and the thermodynamics of the system, graphene offers the appealing opportunity to control thin film morphology through confinement under an atomically-thin cover. Besides,  graphene is also an excellent protective cover against oxidation in air and other environments.

\section*{Acknowledgments}
These experiments were performed at the CIRCE beamline of the ALBA Synchrotron. We thank Jordi Prat for his great technical support.

\section*{References}

\end{document}